\documentclass[traditabstract,letter]{aa} 

\usepackage{graphicx}
\usepackage{txfonts}
\usepackage{array}
\usepackage{amssymb}
\usepackage{natbib}
\usepackage{lscape}
\usepackage{rotating}
\usepackage{longtable}
\newcommand{\pa}{$\pi$\,Aqr}
\newcommand{\xmm}{{\sc{XMM}}\emph{-Newton}}
\begin{document}

\title{\pa\ is another $\gamma$\,Cas object\thanks{Based on observations collected with the ESA science mission \xmm , an ESA Science Mission with instruments and contributions directly funded by ESA Member States and the USA (NASA).}}

\author{Ya\"el~Naz\'e\thanks{F.R.S.-FNRS Research Associate.}
\and Gregor~Rauw
\and Constantin Cazorla
}

\institute{Groupe d'Astrophysique des Hautes Energies, STAR, Universit\'e de Li\`ege, Quartier Agora (B5c, Institut d'Astrophysique et de G\'eophysique), All\'ee du 6 Ao\^ut 19c, B-4000 Sart Tilman, Li\`ege, Belgium\\
\email{naze@astro.ulg.ac.be}
}

\authorrunning{Naz\'e et al.}
\titlerunning{\pa\ in X-rays }
\abstract{ The $\gamma$\,Cas category is a subgroup of Be stars displaying a strong, hard, and variable thermal X-ray emission. An \xmm\ observation of \pa\ reveals spectral and temporal characteristics that clearly make this Be star another member of the $\gamma$\,Cas category. Furthermore, \pa\ is a binary but, contrary to $\gamma$\,Cas, the nature of the companion to the Be star is known; it is a non-degenerate (stellar) object and its small separation from the Be star does not leave much room for a putative {\it compact} object close to the Be disk. This renders the accretion scenario difficult to apply in this system, and, hence, this discovery favors a disk-related origin for the $\gamma$\,Cas phenomenon. }
\maketitle

\section{Introduction}
While its peculiar emission had been known for decades, $\gamma$\,Cas was only recognized in recent years as the prototype of a separate class of X-ray emitters. Currently, there are a dozen Be stars belonging to this category (9 objects + 4 candidates, we refer to \citealt{smi16} for a review). Their X-ray luminosities exceed the faint intrinsic emission of massive stars by an order of magnitude, but are significantly below the bright emission from accreting high-mass X-ray binaries. The $\gamma$\,Cas high-energy emissions appear thermal in nature, though with a very high (10--20\,keV) temperature component, and they are variable on timescales from minutes to years. 

Two classes of scenario have been proposed to explain these properties. The first one considers accretion onto a compact object. This companion is usually considered to be a white dwarf (see e.g., \citealt{mur86}), though the neutron star possibility was recently revived by \citet{pos17}. In the latter case, if the magnetized neutron star spins fast enough, the accretion might be inhibited (the so-called ``propeller'' stage) and the resulting X-ray luminosity remains moderate. The second scenario invokes Be-disk magnetic interactions, with magnetic reconnections giving rise to X-rays \citep{rob00,smi16}. The stellar magnetic field would be localized, arising from subsurface convective layers at the equator, to agree with the overall lack of global, dipolar fields in Be stars \citep{gru12}. Such localized features would also explain the observed UV photometric and spectroscopic variations \citep{smi98}. 

It is in this context that we report in this letter, for the first time, the X-ray properties of the Be star \pa\ (B1V, \citealt{bjo02}), and their resulting constraints on the origin of the $\gamma$\,Cas phenomenon. 

\section{Observations and data reduction} 

\pa\ is a variable Be star rotating half-critically \citep{bjo02,fre05,hua08} and displaying anomalous features in the UV domain \citep{smi06sole}. \xmm\ observed this object in mid-November 2013 for 50\,ks in the framework of our legacy program on B stars (PI Naz\'e, ObsID 0720390701). The data were reduced with SAS (Science Analysis Software) v16.0.0 using calibration files available in Spring 2016 and following the recommendations of the \xmm\ team\footnote{SAS threads, see \\ http://xmm.esac.esa.int/sas/current/documentation/threads/ }. 

After the initial pipeline processing, the EPIC (European Photon Imaging Camera) observations, taken in full-frame mode and with the thick filter (to reject optical/UV light), were filtered to keep only the best-quality data ({\sc{pattern}} 0--12 for MOS and 0--4 for pn). Background flares were detected and times for which the count rate beyond 10\,keV was greater than 0.2\,cts\,s$^{-1}$ for MOS and 0.5\,cts\,s$^{-1}$ for pn were cut before engaging in further analyses. A source detection was performed on each EPIC dataset using the task {\it edetect\_chain} on the 0.4--10.\,keV energy band and for a log-likelihood of 10. The equivalent on-axis, full point spread function (PSF) count rates in this band are 0.726$\pm$0.004, 0.715$\pm$0.004, and 2.206$\pm$0.008\,cts\,s$^{-1}$ for MOS1, MOS2, and pn, respectively. Since those values are slightly over the pile-up limits for the full frame mode, we checked for the presence of pile-up using the task {\it epatplot}, which calculates the distribution of event patterns, and found no obvious signature of pile-up. Nevertheless, all further analyses were performed considering two regions; a circle and an annulus to excise the potentially piled-up core of the PSF.

We extracted EPIC spectra of \pa\ using the task {\it{especget}} in circular regions of 50\arcsec\ radius for MOS and 35\arcsec\ for pn (to avoid CCD gaps) centred on the Simbad position of \pa. We also extracted the spectra using annuli with inner radii fixed to five half instrumental pixel size (i.e., 2.75 and 10.25\arcsec\ for MOS and pn, respectively) to avoid systematic flux inaccuracies$^1$. For the background, a circular region with the same radius as the source was chosen in a region devoid of sources and as close as possible to the target. Dedicated Ancillary Response File (ARF) and Redistribution Matrix File (RMF) response matrices, which are used to calibrate the flux and energy axes, respectively, were also calculated by this task. EPIC spectra were grouped with {\it{specgroup}} to obtain an oversampling factor of five and to ensure that a minimum signal-to-noise ratio of 3 (i.e., a minimum of 10 counts) was reached in each spectral bin of the background-corrected spectra; unreliable bins below 0.25\,keV were discarded. EPIC light curves of \pa\ were extracted for time bins of 5\,s, 10\,s, 100\,s, and 1\,ks in the same regions as the spectra, and in the 0.3--1.0 (soft), 1.0--2.0 (medium), 2.0--10.0 (hard), and 0.3--10.0\,keV (total) energy bands. These were further processed by the task {\it epiclccorr}, which corrects for loss of photons due to  vignetting, off-axis angle, or other problems such as bad pixels. In addition, to avoid very large errors, we discarded bins displaying effective exposure times lower than 50\% of the time bin length. 

Reflection Grating Spectrometer (RGS) data were also processed using the initial pipeline. As for EPIC data, flare filtering was also applied (using a threshold of 0.1\,cts\,s$^{-1}$). The source and background spectra were extracted in the default regions as there is no neighboring object. Dedicated response files were calculated for both orders and both RGS instruments, and were subsequently attached to the source spectra for analysis. In view of the spectral shape, a grouping was performed to obtain at least 20 cts per bin. There is no pile-up in these data.

\section{Spectral characteristics}

\begin{figure}
\includegraphics[width=8.5cm]{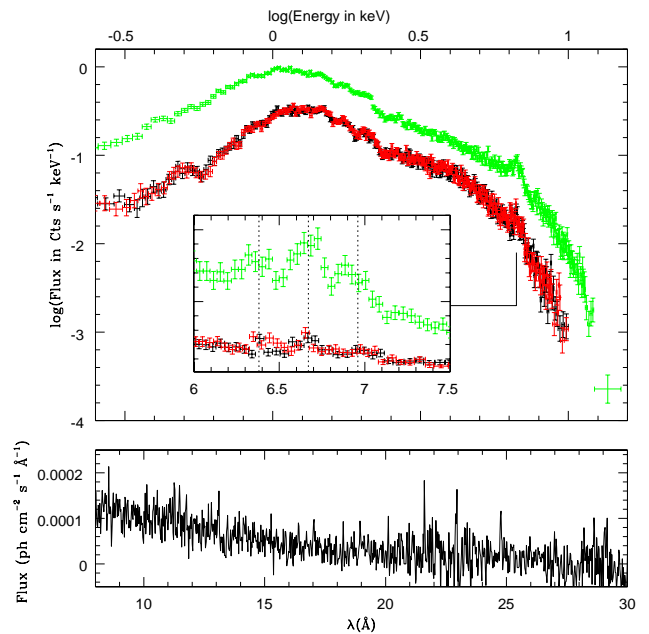}
\caption{The three EPIC spectra (top, MOS1 in black, MOS2 in red, and pn in green) extracted in a circular region centered on \pa\ and the combined RGS spectrum (bottom). A close-up on the FeK$\alpha$ complex, with the positions of the three components shown by dotted lines, is included in the top panel. The EPIC data extracted in an annular source region (see Sect. 2) are similar, but noisier.}
\label{spec}
\end{figure}

The X-ray spectrum of \pa\ is shown in Fig. \ref{spec}. The high-resolution RGS spectra appear rather featureless, though with a detectable slope - there is only marginal evidence for the presence of N\,{\sc vii}$\lambda$24.78\AA. If lines exist, they are buried in the noise. The lower-resolution EPIC spectra, however, reveal the lines of the FeK$\alpha$ complex - the iron fluorescence line at 6.38\,keV, and the two ionized iron lines at 6.67 and 6.96\,keV. The presence of the latter two lines clearly indicates that the plasma is thermal, and we therefore fitted the spectra within Xspec v12.9.0i using an absorbed $apec$ complemented by a Gaussian line to fit the fluorescence feature. Table \ref{fit} provides these fitting results for both extraction regions. It shows that, as expected from {\it epatplot} results, pile-up is very limited since the derived parameters are similar for both extraction regions (with/without the core). Nevertheless, we also attempted a fit with a power law, with an additional Gaussian line which fitted the 6.67\,keV line (see Table \ref{fit}). 

The X-ray flux in the 0.5--10.keV energy band, corrected for interstellar absorption, can be compared to the optical flux. To this aim, we first use the distance and bolometric luminosities of \citet[$d$=340\,pc, $\log(L_{\rm BOL}/L_{\odot})$=4.1]{bjo02}, which yields $L_{\rm X}=1.5\times10^{32}$\,erg\,s$^{-1}$ in 0.5--10.\,keV and $\log(L_{\rm X}/L_{\rm BOL})=-5.5$. Choosing instead the values of \citet[$d$=740\,pc, $\log(L_{\rm BOL}/L_{\odot})$=4.7]{zha13} yields $L_{\rm X}=7.0\times10^{32}$\,erg\,s$^{-1}$ and $\log(L_{\rm X}/L_{\rm BOL})=-5.4$. The value of the ratio  is clearly larger than the value of $\log(L_{\rm X}/L_{\rm BOL})\sim -7$ typical in this energy band for ``normal'' O and very early B stars, knowing that this ratio is often lower for B1 stars \citep{naz11,rau15}. It is also larger than expected for massive stars with strong, large-scale (dipolar) magnetic fields \citep{naz14}, though this could appear irrelevant in this context since Be stars are known to not host such fields \citep[see, e.g.,][]{gru12}. In addition, the X-ray luminosity is too large for contamination by a (putative) pre-main-sequence (PMS) companion, and too small for X-ray binaries. However, a luminosity of $10^{32-33}$\,erg\,s$^{-1}$ is quite typical of the subclass of Be stars called $\gamma$\,Cas objects \citep[see the review of][and references therein]{smi16}. 

In the context of $\gamma$\,Cas objects, the X-ray emission of \pa\ appears much more typical, and not only because of the flux. Indeed, their spectrum is characterized by the presence of a very hot plasma, around 10--20\,keV, as we observe. This hot component is often accompanied by lower temperature plasma \citep{smi04,lop06,lop10,tor12}, but not always \citep{rau13}. Also, iron abundance is found to be subsolar for such objects \citep[e.g.,][]{lop07}. Therefore, we did additional spectral fittings, allowing for non-solar iron abundances and considering either one or two thermal components (see Table \ref{fit}): the fit quality slightly improves, and results further underline that \pa\ behaves like other $\gamma$\,Cas objects. In contrast, such a high temperature is not present in ``normal'' or magnetic OB stars. Furthermore, the absorbing column needed in addition to the interstellar one, which represents local absorption, is also similar to those derived for other $\gamma$\,Cas objects. 

\begin{table*}
\centering
\caption{Results of the spectral fits.  }
\label{fit}
\begin{tabular}{lccccccc}
\hline\hline
\multicolumn{8}{l}{\tiny Models $tbabs_{ism}*phabs*(apec+gauss)$, or $tbabs_{ism}*vphabs*(\sum vapec+gauss)$ for $^*$ and $^{**}$, $tbabs_{ism}*(vphabs_1*vapec_1+vphabs_2*(vapec_2+gauss))$ for $^{\dagger}$}\\
    & $N_{\rm H}$ & $kT$ & $norm$ & $center_{line}$ & $strength_{line}$ & $\chi^2$ (dof) & $F^{\rm obs}_{\rm X}$ \\
    & (10$^{22}$\,cm$^{-2}$) & (keV) & ($10^{-3}$\,cm$^{-5}$) & (keV) & ($10^{-6}$\,ph\,cm$^{-2}$\,s$^{-1}$)& & ($10^{-11}$\,erg\,cm$^{-2}$\,s$^{-1}$)\\
\hline
EPIC, C     & 0.215$\pm$0.002 & 11.7$\pm$0.2 & 6.32$\pm$0.02 & 6.371$\pm$0.019 & 5.35$\pm$0.86 & 1.23 (804) & 1.053$\pm$0.004 \\
EPIC, A     & 0.223$\pm$0.003 & 10.4$\pm$0.2 & 6.41$\pm$0.02 & 6.355$\pm$0.020 & 5.76$\pm$1.24 & 1.20 (754) & 1.061$\pm$0.005 \\
EPIC+RGS, A & 0.225$\pm$0.003 & 10.5$\pm$0.2 & 6.39$\pm$0.02 & 6.355$\pm$0.022 & 5.84$\pm$1.18 & 1.20 (1218) & 1.057$\pm$0.005 \\
EPIC+RGS, A$^*$    & 0.235$\pm$0.004 & 10.5$\pm$0.3 & 6.48$\pm$0.02 & 6.374$\pm$0.022 & 6.19$\pm$1.19 & 1.15 (1217) & 1.042$\pm$0.006 \\
EPIC+RGS, A$^{**}$ & 0.238$\pm$0.004 & 1.41$\pm$0.18& 0.19$\pm$0.07 & 6.372$\pm$0.021 & 5.87$\pm$1.19 & 1.12 (1215) & 1.050$\pm$0.007 \\
                  &                 & 12.0$\pm$0.5 & 6.36$\pm$0.05 & \\
EPIC+RGS, A$^{\dagger}$ & 0.$\pm$0.03     & 1.44$\pm$0.13& 0.17$\pm$0.06 &                &               & 1.11 (1214) & 1.048$\pm$0.006 \\
                       & 0.263$\pm$0.010 & 11.6$\pm$0.5 & 6.40$\pm$0.04 &6.370$\pm$0.015 & 5.89$\pm$1.18 \\
\hline
\multicolumn{6}{l}{Model $tbabs_{ism}*phabs*(pow+gauss)$}\\
    & $N_{\rm H}$ & $\Gamma$ & $norm$ & $center_{line}$ & $strength_{line}$ & $\chi^2$ (dof) & $F^{\rm obs}_{\rm X}$ \\
    & (10$^{22}$\,cm$^{-2}$) & & ($10^{-3}$\,cm$^{-5}$) & (keV) & ($10^{-5}$\,ph\,cm$^{-2}$\,s$^{-1}$) & & ($10^{-11}$\,erg\,cm$^{-2}$\,s$^{-1}$)\\
\hline
EPIC, C     & 0.287$\pm$0.004 & 1.591$\pm$0.007 & 1.717$\pm$0.014 & 6.680$\pm$0.005 & 1.25$\pm$0.10 & 1.32 (804) & 1.054$\pm$0.004 \\
EPIC, A     & 0.302$\pm$0.003 & 1.618$\pm$0.006 & 1.806$\pm$0.011 & 6.685$\pm$0.023 & 1.20$\pm$0.14 & 1.29 (754) & 1.064$\pm$0.006 \\
EPIC+RGS, A & 0.300$\pm$0.004 & 1.608$\pm$0.007 & 1.778$\pm$0.014 & 6.676$\pm$0.009 & 1.19$\pm$0.14 & 1.28 (1218) & 1.062$\pm$0.006 \\
\hline
\end{tabular}
\\
\tablefoot{``EPIC'' indicates a simultaneous fit of all EPIC spectra, ``EPIC+RGS'' a simultaneous fit of all \xmm\ spectra; and ``C'' and ``A'' indicate EPIC extractions in a circle or annulus (see Sect. 2). The interstellar column was fixed to $3.16\times 10^{20}$\,cm$^{-2}$ \citep{gud12} and the Gaussian width to zero since these lines are not resolved by EPIC cameras. Solar abundances of \citet{asp09} were used except for cases marked by $^*$, $^{**}$, or $^{\dagger}$: in these cases, $vapec$ and $vphabs$ were used and the best-fit abundances of iron in number, relative to hydrogen and relative to the solar value amount to $A_{Fe}^*=0.62\pm0.05$, $A_{Fe}^{**}=0.71\pm0.05$, and $A_{Fe}^{\dagger}=0.67\pm0.05, $ respectively. Fluxes, expressed in the 0.5--10.0\,keV band correspond to observed ones. Errors (found using the ``error'' command for the spectral parameters and the ``flux err'' command for the fluxes) correspond to 1$\sigma$; whenever errors were asymmetric, the highest value is provided here.  }
\end{table*}

\section{Temporal variability}

\begin{figure*}
\includegraphics[width=6cm]{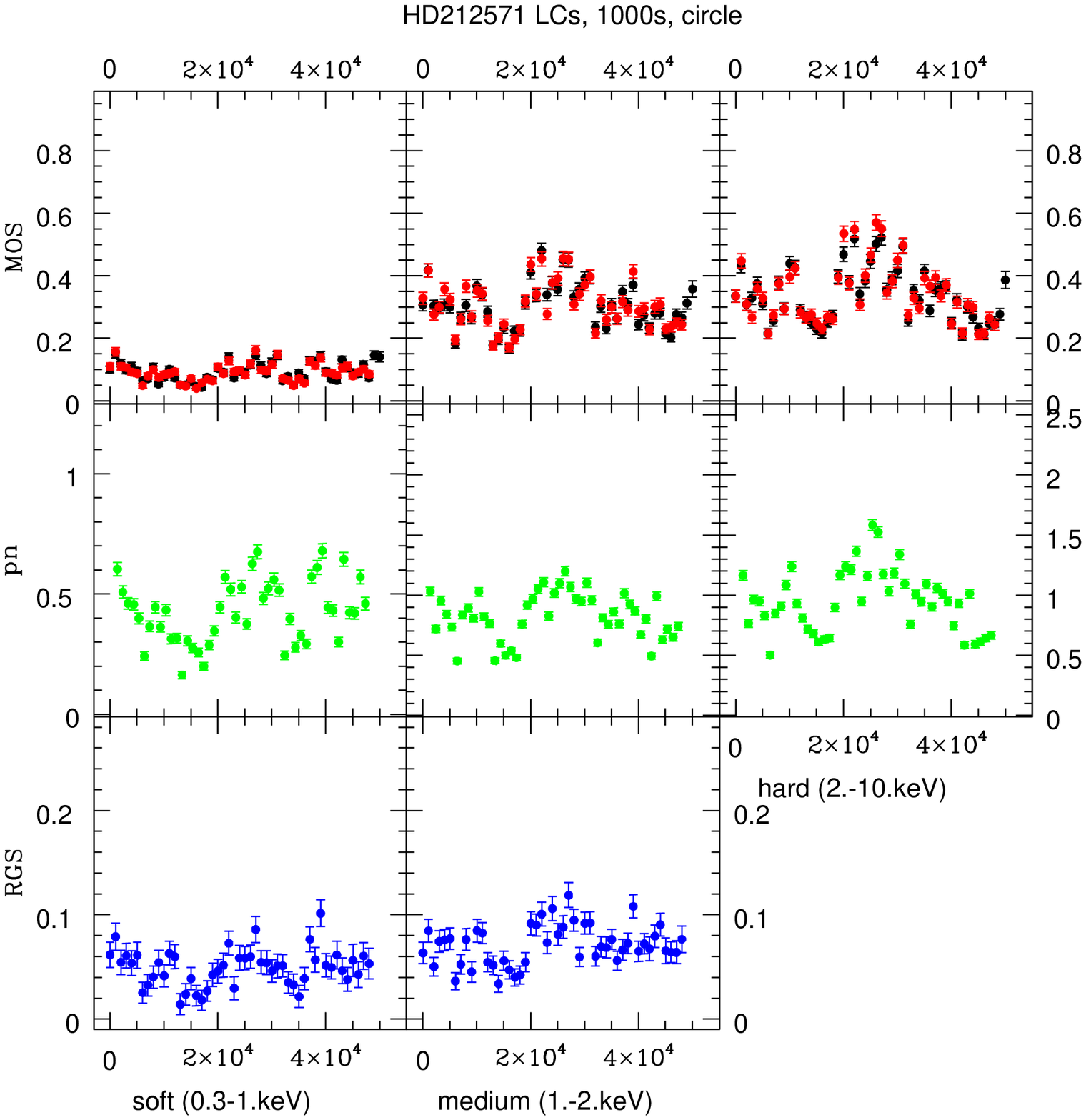}
\includegraphics[width=6cm]{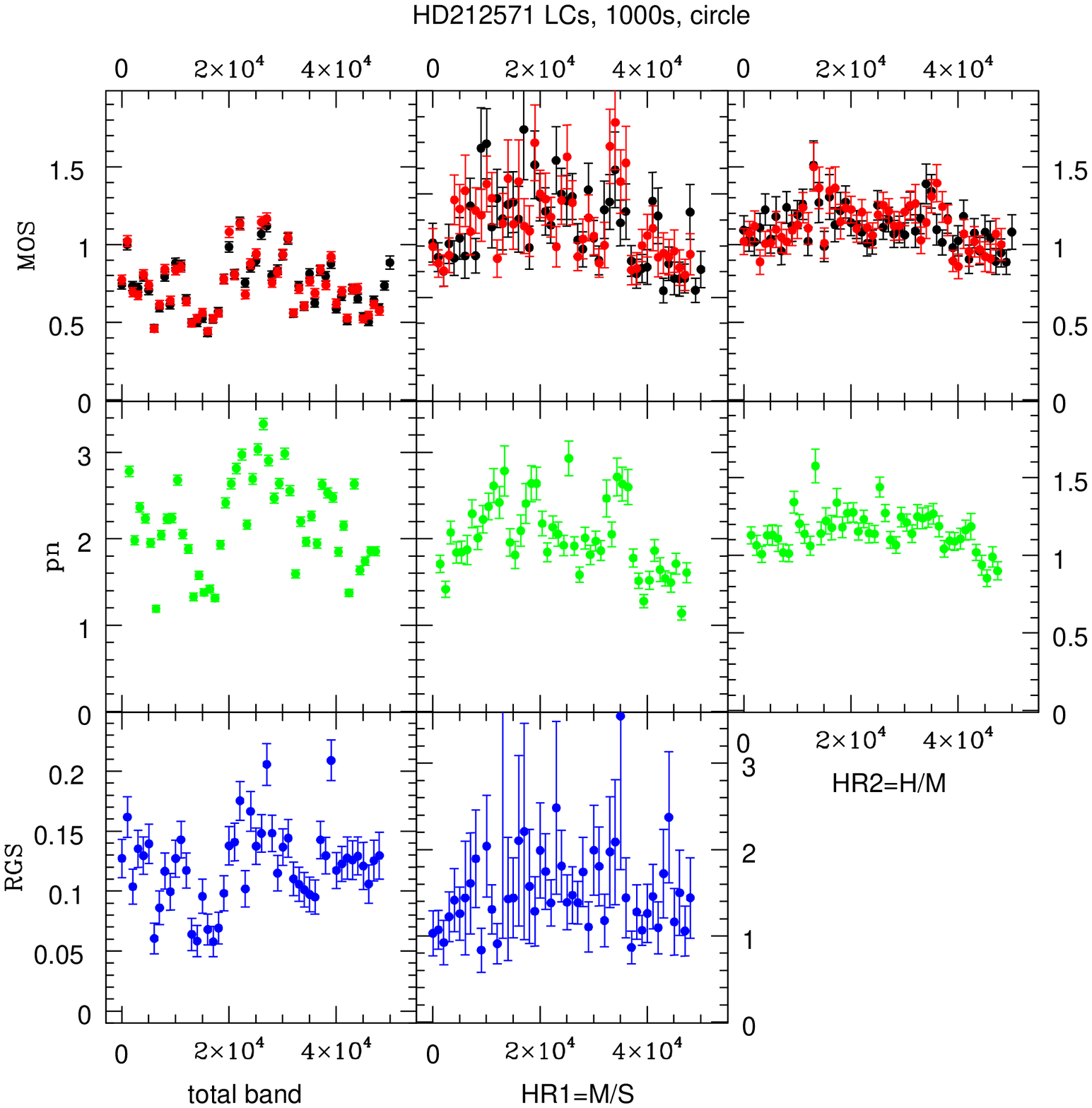}
\includegraphics[width=6cm]{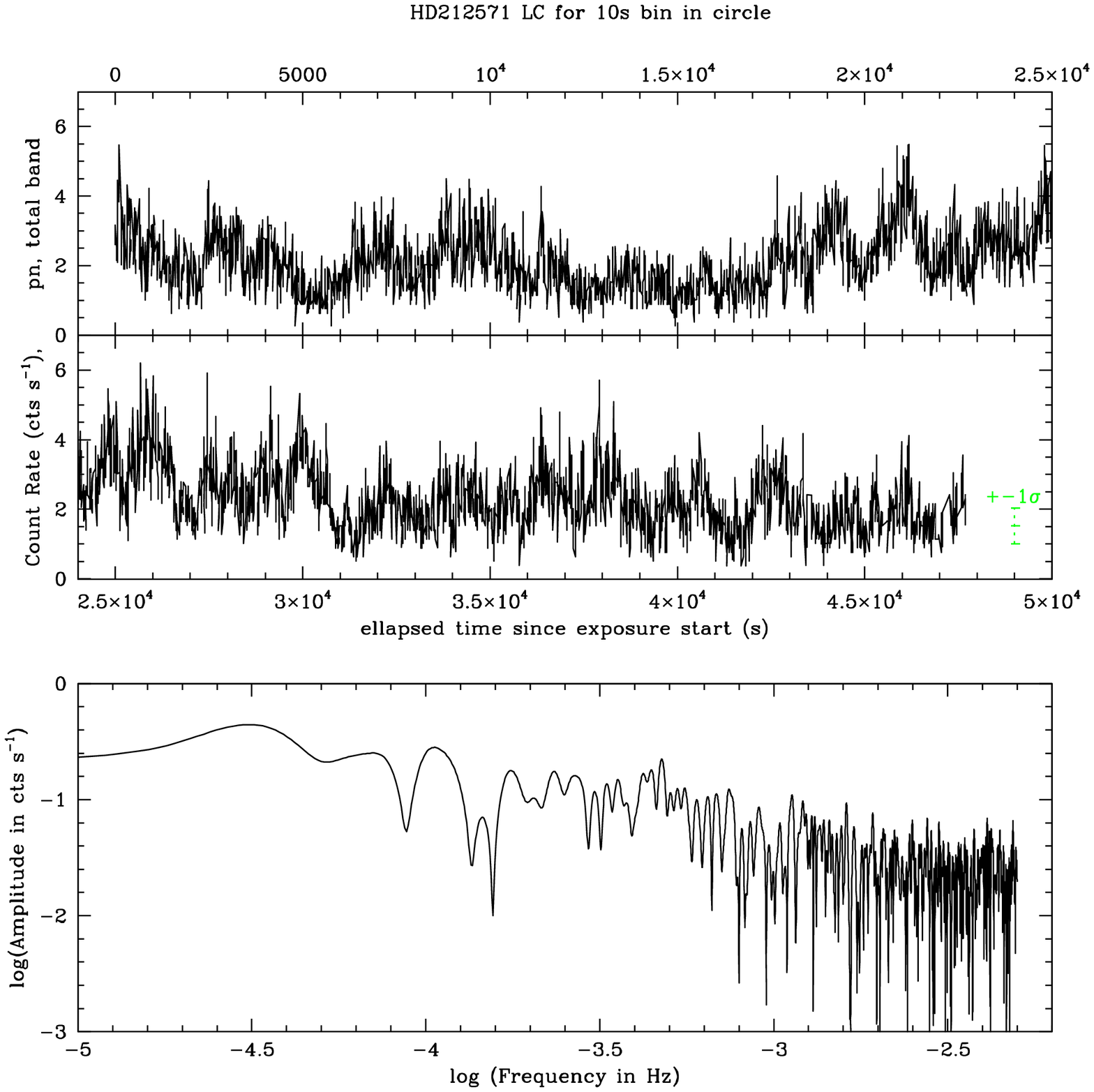}
\caption{{\it Left and middle:} Background-subtracted lightcurves, with a 1ks bin and for different energy bands, recorded with the three EPIC cameras and with the RGS (MOS1 in black, MOS2 in red, pn in green, combination of both RGS and both orders in blue) extracted in a circular region around \pa. The EPIC data extracted in an annular source region (see Sect. 2) are similar, but noisier. {\it Right:} Background-subtracted lightcurve with a 10s bin recorded for pn, along with its Fourier periodogram \citep{hmm}. The green dotted line indicates the typical $\pm1\sigma$ error around each data point. The short-term variations clearly are larger than this, demonstrating the presence of flares.}
\label{lc}
\end{figure*}

The evolution of the X-ray emission of \pa\ during the \xmm\ exposure is shown in Fig. \ref{lc}. As can be seen, large variations in flux are detected, of a factor of $\sim3$ when considering 1ks bins. These changes in flux are not accompanied by changes in hardness: the hardness ratios only show limited variations. The variations occur on both short and long timescales: there are short flares superimposed on slower variations. Such a light curve
is clearly reminiscent of what was detected with \xmm, {\it Chandra}, or {\it RXTE} for other $\gamma$\,Cas objects \citep{smi06,lop10,smi12,tor13,shr15} but again different from what is observed in ``normal'' or magnetic OB stars \citep[e.g.,][]{naz13,naz14}. Analyzing the light curves with period search methods \citep{hmm,gra13} yields no obvious periodicity, marking the absence of pulsations. The periodogram tends to increase at low frequencies, however this simply reflects the presence of slow variations; those timescales do not yield a coherent folded light curve. In log-log scale (Fig. \ref{lc}), the periodogram appears similar to that found for HD\,110432 with {\it Chandra} \citep{tor12}; it does not show a clear $1/f$ trend, as detected notably in $\gamma$\,Cas by \citet{lop10}, probably because of the larger noise in our data. This noise also prevents us from significantly detecting very short events associated with increased absorption \citep{ham16}.

Longer-term variations may also exist. Unfortunately, \pa\ was observed only twice previously in the X-ray domain, both with short exposures. The first observation (with 275\,s duration) was taken in the framework of the {\it ROSAT} All-Sky Survey (RASS). \pa\ was reported under the name of 1RXS J222517.0+012238 with a count rate of 0.124$\pm$0.023\,cts\,s$^{-1}$ \citep{vog99}. This rather high count rate, which added to the peculiar character of \pa, was remarked by \citet{smi06sole}. The hardness ratio, computed from $(B-A)/(B+A)$ where A and B are the count rates in the 0.5--0.9 and 0.9--2.0\,keV domains, respectively, amounted to 0.48$\pm$0.17. Though the ratio is not well constrained in view of its large error, it is certainly positive, indicating the source to be rather hard, in agreement with what we observed. We folded our best-fit thermal model (Table \ref{fit}) through the response of {\it ROSAT} (for gain=1 and the PSPCB camera, as adequate for the RASS), and found a count rate of 0.188\,cts\,s$^{-1}$ in the 0.1--2.0\,keV energy band. The brightening between 1990--1 and 2013 thus amounts to $\sim$50\%, but it should be noted that both values are compatible at the 3$\sigma$ level. The second X-ray observation (about 2s) of \pa\ occurred in May 2004 during a slew of \xmm. \pa\ was then cataloged as XMMSL1 J222517.0+012223 with a combined EPIC count rate of 2.4$\pm$0.7\,cts\,s$^{-1}$ in the 0.2--12.\,keV range \citep{sax08} or as XMMSL2\footnote{This second version of the slew survey catalog is accessible through the XSA archive at https://www.cosmos.esa.int/web/xmm-newton} J222517.0+012226 with a count rate of 2.8$\pm$1.1\,cts\,s$^{-1}$. In contrast, the same \xmm\ team used our observation to derive a count rate in the 3XMM catalog (where \pa\ appears as 3XMM J222516.6+012238) of 3.73$\pm$0.01\,cts\,s$^{-1}$ in the same band and for the same instrument combination as for the slew survey. Again, a $\sim$50\% brightening is detected, but this difference is only at the 1--2$\sigma$ level. However, it may be noted that brightenings were detected in other $\gamma$\,Cas objects with similar amplitude (e.g., $L_{\rm X}=3.5-8.8\times10^{32}$\,erg\,cm$^{-2}$\,s$^{-1}$ for HD\,110432 using RXTE in \citealt{smi12}) or similar timescales (e.g., years for $\gamma$\,Cas in \citealt{mot15}), while they are not observed in ``normal'' OB stars. The detection of long-term changes actually provides further support for the identification of \pa\ as a $\gamma$\,Cas object. 

\section{Discussion and Conclusion}

Our \xmm\ observation of \pa\ clearly reveals it to belong to the class of $\gamma$\,Cas objects; it has a hard, thermal X-ray spectrum ($kT\sim10.5$\,keV) with moderate absorption ($N_{\rm H}\sim2\times10^{21}$\,cm$^{-2}$) and moderate brightness ($\log(L_{\rm X}/L_{\rm BOL})\sim-5.5$). The iron fluorescence line is detected, further indicating the presence of cooler circumstellar material as expected around a Be star. Finally, as usual for $\gamma$\,Cas objects, \pa\ displays flux variations on short (minutes), medium (hours), and long (years) timescales. 

Could this conclusion be misled? Let us examine the stellar properties of \pa. The star is known to display large variations of its emission lines. A very active phase occurred from the 1950s to 1995 but the emissions have been much smaller since then; though they are still somewhat variable (see Fig. 1 of \citealt{bjo02} and Fig. 3 of \citealt{zha13}). The BeSS\footnote{http://basebe.obspm.fr/basebe/} database provides spectra obtained on 15 and 19 November 2013, bracketing our X-ray exposure date. Both H$\alpha$ profiles are similar and in agreement with the known properties of the star (the violet emission peak is slightly stronger than the red one, as expected for that date, and the $EW$ is $\sim3$\,\AA, compatible with the ``low'' state - see \citealt{zha13} for comparison). \pa\ was thus not in a peculiar state at the time of the \xmm\ observation. 

The star is also known to be a binary. Indeed, \citet{bjo02} detected sinusoidal motions, with $P$=84.1\,d, of both the emission and absorption components of the H$\alpha$ line. While the primary has 10--14\,$M_{\odot}$, the companion was found to have a mass of 2--3\,$M_{\odot}$, and \citet{bjo02} consider it to be an A or F-type star of the main sequence. At high energies, A-type stars are known to be X-ray faint though short, intense flares may occur in strongly magnetic cases but with a different light curve (a short, localized, and single flare), less extreme temperatures, and less extreme luminosities than observed here \citep[and references therein]{rob14}. F-type stars, even if on the PMS, display coronal X-ray emissions with levels $<10^{31}$\,erg\,s$^{-1}$ \citep{gud09}. Such stars cannot account for the observed X-rays, which are thus truly associated with the Be star.

As recalled in the introduction, two main models currently compete to explain the $\gamma$\,Cas-like behavior in X-rays: Accretion onto a compact object \citep[and references therein]{pos17} and magnetic interactions in the disk \citep{mot15,smi16}. In this context, it must be noted that the companion on the circular, 84.1\,d orbit is certainly {not} a compact object \citep{bjo02,zha13}. Indeed, the large mass of the companion is incompatible with masses of white dwarves and neutron stars \citep{oze16}. Furthermore, the orbital configuration of \pa\ provides strong constraints: The companion of \pa\ is located at $a\sin(i)$=0.96\,AU \citep[with an inclination $\sim70^{\circ}$]{bjo02} and the circumstellar disk has a radius of 65\,$R_{\odot}$ \citep{zha13}, or one third of the separation. With such a geometry, there is not much room in the system to place a compact object relatively close to the Be star, and a putative compact companion in stable orbit would have to be quite distant from the Be+AF pair, rendering wind accretion much less efficient than required. The detection of a $\gamma$\,Cas-like behavior in the stellar binary \pa\ therefore seems to favor the magnetic interaction model. 

Further investigation is needed, though, as the current dataset only probes a single epoch. The X-ray emission of \pa\ should now be monitored, along with its optical characteristics, on several timescales; in particular the 84.1\,d orbital period (since disk variations occur with this period; \citealt{zha13}) but also longer timescales to probe outbursts. Only then will the exact relationship between the disk and the X-ray emission be unveiled, and the $\gamma$\,Cas phenomenon fully understood.

\begin{acknowledgements}
We acknowledge support from the Fonds National de la Recherche Scientifique (Belgium), the Communaut\'e Fran\c caise de Belgique, the PRODEX \xmm\ contract, and an ARC grant for concerted research actions financed by the French community of Belgium (Wallonia-Brussels Federation). ADS and CDS were used in  preparing this document. 
\end{acknowledgements}


\begin{thebibliography}{00}
\bibitem[Asplund et al.(2009)]{asp09} Asplund, M., Grevesse, N., Sauval, A.J., \& Scott, P.\ 2009, \araa, 47, 481 
\bibitem[Bjorkman et al.(2002)]{bjo02} Bjorkman, K.~S., Miroshnichenko, A.~S., McDavid, D., \& Pogrosheva, T.~M.\ 2002, \apj, 573, 812 
\bibitem[Fr{\'e}mat et al.(2005)]{fre05} Fr{\'e}mat, Y., Zorec, J., Hubert, A.-M., \& Floquet, M.\ 2005, \aap, 440, 305 
\bibitem[Graham et al.(2013)]{gra13} Graham, M.~J., Drake, A.~J., Djorgovski, S.~G., Mahabal, A.~A., \& Donalek, C.\ 2013, \mnras, 434, 2629 
\bibitem[Grunhut et al.(2012)]{gru12} Grunhut, J.~H., Wade, G.~A., \& MiMeS Collaboration 2012, American Institute of Physics Conference Series, 1429, 67 
\bibitem[Gudennavar et al.(2012)]{gud12} Gudennavar, S.~B., Bubbly, S.~G., Preethi, K., \& Murthy, J.\ 2012, \apjs, 199, 8 
\bibitem[G{\"u}del \& Naz{\'e}(2009)]{gud09} G{\"u}del, M., \& Naz{\'e}, Y.\ 2009, \aapr, 17, 309 
\bibitem[Hamaguchi et al.(2016)]{ham16} Hamaguchi, K., Oskinova, L., Russell, C.~M.~P., et al.\ 2016, \apj, 832, 140 
\bibitem[Heck et al.(1985)]{hmm} Heck, A., Manfroid, J., \& Mersch, G.\ 1985, \aaps, 59, 63 
\bibitem[Huang \& Gies(2008)]{hua08} Huang, W., \& Gies, D.~R.\ 2008, \apj, 683, 1045-1051 
\bibitem[Lopes de Oliveira et al.(2006)]{lop06} Lopes de Oliveira, R., Motch, C., Haberl, F., Negueruela, I., \& Janot-Pacheco, E.\ 2006, \aap, 454, 265 
\bibitem[Lopes de Oliveira et al.(2007)]{lop07} Lopes de Oliveira, R., Motch, C., Smith, M.~A., Negueruela, I., \& Torrej{\'o}n, J.~M.\ 2007, \aap, 474, 983 
\bibitem[Lopes de Oliveira et al.(2010)]{lop10} Lopes de Oliveira, R., Smith, M.~A., \& Motch, C.\ 2010, \aap, 512, A22 
\bibitem[Motch et al.(2015)]{mot15} Motch, C., Lopes de Oliveira, R., \& Smith, M.~A.\ 2015, \apj, 806, 177 
\bibitem[Murakami et al.(1986)]{mur86} Murakami, T., Koyama, K., Inoue, H., \& Agrawal, P.~C.\ 1986, \apjl, 310, L31 
\bibitem[Naz{\'e} et al.(2011)]{naz11} Naz{\'e}, Y., Broos, P.~S., Oskinova, L., et al.\ 2011, \apjs, 194, 7 
\bibitem[Naz{\'e} et al.(2013)]{naz13} Naz{\'e}, Y., Oskinova, L.~M., \& Gosset, E.\ 2013, \apj, 763, 143 
\bibitem[Naz{\'e} et al.(2014)]{naz14} Naz{\'e}, Y., Petit, V., Rinbrand, M., et al.\ 2014, \apjs, 215, 10 (erratum 2016, ApJS, 224, 13)
\bibitem[{\"O}zel \& Freire(2016)]{oze16} {\"O}zel, F., \& Freire, P.\ 2016, \araa, 54, 401 
\bibitem[Postnov et al.(2017)]{pos17} Postnov, K., Oskinova, L., \& Torrej{\'o}n, J.~M.\ 2017, \mnras, 465, L119 
\bibitem[Rauw et al.(2013)]{rau13} Rauw, G., Naz{\'e}, Y., Spano, M., Morel, T., \& ud-Doula, A.\ 2013, \aap, 555, L9 
\bibitem[Rauw et al.(2015)]{rau15} Rauw, G., Naz{\'e}, Y., Wright, N.~J., et al.\ 2015, \apjs, 221, 1 
\bibitem[Robinson \& Smith(2000)]{rob00} Robinson, R.~D., \& Smith, M.~A.\ 2000, \apj, 540, 474 
\bibitem[Robrade(2014)]{rob14} Robrade, J.\ 2014, Putting A Stars into Context: Evolution, Environment, and Related Stars, 425 
\bibitem[Saxton et al.(2008)]{sax08} Saxton, R.~D., Read, A.~M., Esquej, P., et al.\ 2008, \aap, 480, 611 
\bibitem[Shrader et al.(2015)]{shr15} Shrader, C.~R., Hamaguchi, K., Sturner, S.~J., et al.\ 2015, \apj, 799, 84 
\bibitem[Smith et al.(1998)]{smi98} Smith, M.~A., Robinson, R.~D., \& Hatzes, A.~P.\ 1998, \apj, 507, 945 
\bibitem[Smith et al.(2004)]{smi04} Smith, M.~A., Cohen, D.~H., Gu, M.~F., et al.\ 2004, \apj, 600, 972 
\bibitem[Smith(2006)]{smi06sole} Smith, M.~A.\ 2006, \aap, 459, 215 
\bibitem[Smith et al.(2006)]{smi06} Smith, M.~A., Henry, G.~W., \& Vishniac, E.\ 2006, \apj, 647, 1375 
\bibitem[Smith et al.(2012)]{smi12} Smith, M.~A., Lopes de Oliveira, R., \& Motch, C.\ 2012, \apj, 755, 64 
\bibitem[Smith et al.(2016)]{smi16} Smith, M.~A., Lopes de Oliveira, R., \& Motch, C.\ 2016, Advances in Space Research, 58, 782 
\bibitem[Torrej{\'o}n et al.(2012)]{tor12} Torrej{\'o}n, J.~M., Schulz, N.~S., \& Nowak, M.~A.\ 2012, \apj, 750, 75 
\bibitem[Torrej{\'o}n et al.(2013)]{tor13} Torrej{\'o}n, J.~M., Schulz, N.~S., Nowak, M.~A., Testa, P., \& Rodes, J.~J.\ 2013, \apj, 765, 13 
\bibitem[Voges et al.(1999)]{vog99} Voges, W., Aschenbach, B., Boller, T., et al.\ 1999, \aap, 349, 389 
\bibitem[Zharikov et al.(2013)]{zha13} Zharikov, S.~V., Miroshnichenko, A.~S., Pollmann, E., et al.\ 2013, \aap, 560, A30 


\end{thebibliography}
\end{document}